\documentclass[11pt,a4paper]{article}
\usepackage{amsmath, amsfonts, amssymb}
\usepackage{a4wide}
\usepackage{parskip}
\usepackage{enumitem}
\usepackage{xcolor}

\usepackage{hyperref}
\usepackage{booktabs}
\usepackage{caption}
\usepackage{siunitx}
\usepackage{tabularx}
\usepackage{graphicx}
\usepackage{adjustbox}
\usepackage{multirow}
\usepackage{amsthm}
\usepackage[noadjust]{cite}
\def\proof{\noindent{\textit{Proof. }}}
\def\qed{\hfill {$\square$}\goodbreak \medskip}

\newtheorem{theorem}{Theorem}[section]
\newtheorem{lemma}[theorem]{Lemma}
\theoremstyle{definition}
\newtheorem{definition}[theorem]{Definition}
\newtheorem{example}[theorem]{Example}
\theoremstyle{conjecture}

\theoremstyle{proposition}
\newtheorem{proposition}[theorem]{Proposition}
\theoremstyle{remark}
\newtheorem{remark}[theorem]{Remark}

\theoremstyle{corollary}

\numberwithin{equation}{section}

\usepackage{lipsum} 

\usepackage{tikz,xcolor,hyperref}

\definecolor{lime}{HTML}{A6CE39}
\DeclareRobustCommand{\orcidicon}{%
	\begin{tikzpicture}
		\draw[lime, fill=lime] (0,0) 
		circle [radius=0.16] 
		node[white] {{\fontfamily{qag}\selectfont \tiny ID}};
		\draw[white, fill=white] (-0.0625,0.095) 
		circle [radius=0.007];
	\end{tikzpicture}
	\hspace{-2mm}
}

\foreach \x in {A, ..., Z}{%
	\expandafter\xdef\csname orcid\x\endcsname{\noexpand\href{https://orcid.org/\csname orcidauthor\x\endcsname}{\noexpand\orcidicon}}
}

\begin{document}
	\date{}
	{\vspace{0.01in}
		\title{Minimal and Optimal binary codes obtained using $C_D$-construction over the non-unital ring $I$}
		\author{{\bf Vidya Sagar\footnote{email: {\tt vsagariitd@gmail.com}}\orcidA{}  \;and  \bf Ritumoni Sarma\footnote{email: {\tt ritumoni407@gmail.com}}\orcidB{}} \\ Department of Mathematics,\\ Indian Institute of Technology Delhi,\\Hauz Khas, New Delhi-110016, India. }
		\maketitle
		\begin{abstract}
			In this article, we construct linear codes over the commutative non-unital ring $I$ of size four. We obtain their Lee-weight distributions and study their binary Gray images. Under certain mild conditions, these classes of binary codes are minimal and self-orthogonal. All codes in this article are few-weight codes. Besides, an infinite class of these binary codes consists of distance optimal codes with respect to the Griesmer bound.

			\medskip
			
			
			\medskip
			
			
		\end{abstract}
		\section{INTRODUCTION}		
		Codes over rings of size four, namely, $\mathbb{F}_2\times \mathbb{F}_2$ (in \cite{F2xF2}), $\mathbb{F}_2 + u\mathbb{F}_2$ (in \cite{F2uF2}), $\mathbb{Z}_4$ (in \cite{Z4}) and $\mathbb{F}_4$ (in \cite{F4}), have got huge attention in the area of algebraic coding theory in the past. These are only commutative unital rings among the 11 rings of size four classified by Fine in \cite{Fine}. Out of the 11 rings, $5$ of them are commutative non-unital (denoted as $B$, $C$, $H$, $I$ and $J$) and $2$ of them are non-commutative non-unital (denoted as $E$ and $F$).\par 
		In 2021, Alahmadi et. al. \cite{AlahmadiI1} studied algebraic structure of linear codes over the ring $I$ (see \cite{Fine}) by introducing the notion of QSD codes. After that the authors in \cite{AlahmadiI2} studied a construction of self-orthogonal codes over $I$, and classified Type IV codes and quasi Type IV codes over $I$ up to length $n=6$. Motivated by these works, Kim and Roe in \cite{KimI} further investigated QSD codes, Type IV codes and quasi Type IV codes over $I$ and extended the previous work for lengths $n=7$ and $8$. Since then codes over non-unital rings have been studied by many researchers (for instance, \cite{AlahmadiE2, AlahmadiE1, DNAnonunital, MinjaNonunital, LCDnonUnital}).\par
		
		In this article, we study the linear $I$-code $C_D = \{(v\cdot d)_{d\in D} : v\in I^m\}$, where $D\subseteq I^m$ is an ordered finite multiset, called \textit{the defining set} for $C_D$. In \cite{Ding}, the authors introduced the construction of $C_D$ in order to generalize the Mattson-Solomon transform for cyclic codes. Study of $C_D$ becomes convenient, if the defining set $D$ is constructed via simplicial complexes. In fact, several interesting linear codes have been constructed in the recent past by using simplicial complexes (for instance, \cite{Hyun_Kim, Hyun_Lee, Sagar_Sarma2,shi_x, shi_nonchain,  shi_x2, Shi_guan, shi_qian, shi_xuan, mixed2, wu_zhu, Zhu_Wei}). It is expected that for properly chosen finite fields (more generally, rings) and defining sets, we may be able to discover codes with good parameters using the above construction.\par 
		Yansheng et. al. in \cite{Wu_Li} studied linear codes over $\mathbb{F}_{4}$ and their binary subfield codes, and could produce two infinite families of optimal linear codes. Motivated by this work, the authors in \cite{Sagar_Sarma} studied octanary linear codes using simplicial complexes, and could obtain minimal and optimal codes. After that the authors in \cite{GeneralCase} generalized the work of \cite{Sagar_Sarma} for finite fields of characteristic $2$ by using LFSR sequences. Recently, the authors in \cite{Sagar_Sarma4} studied algebraic structure of linear codes over the ring $E$ (see \cite{Fine}) using simplicial complexes for the first time and obtained several classes of minimal codes, self-orthogonal codes and distance optimal codes.\par		
		The weight distribution (see \cite{wchuffman}) of a linear code contains crucial information regarding error detecting as well as error correcting capacity of the code, and it allows the computation of the error probability of error detection and correction with respect to certain algorithms \cite{Klove}. Few-weight codes are useful because of their connection with strongly regular graphs, difference sets and finite geometry \cite{Few1, Few2}. The minimum Hamming distance of linear codes are well known for their importance in determining error-correcting capacity. As a result, finding optimal linear codes has become one of the central topics of research. In \cite{Hyun_Lee}, the authors showed how optimal codes can be utilized for the construction of secret sharing schemes with nice access structures following the framework discussed in \cite{YuanDing}. Minimal codes are useful to construct the access structure of the secret sharing schemes \cite{Ding_Ding, MasseyMinimal}. This special class of codes is also important as these can be decoded by using the minimum distance decoding rule \cite{suffConMinimalcode}. Normally, it is difficult to identify all the minimal codewords of a given code even over a finite field of characteristic $2$. In view of this, researchers began to investigate minimal codes. Self-orthogonal codes are also essential class of codes. For instance, they can be utilized for the construction of quantum error-correcting codes \cite{Ashikhmin_QECC, Calder_QECC}. For the construction of self-orthogonal codes we refer to \cite{Du_Sihem, Gan_Sihme, Wu_Lee, Wu_Lee_2, Zhou_Li}. \par 
		Inspired by the work of \cite{Wu_Li}, a natural endeavour is to explore the construction of linear codes over non-unital rings with the help of simplicial complexes and investigate the structure of these codes. We consider the commutative non-unital ring $I$ according to the notation of Fine \cite{Fine}. We, in this article, choose a defining set $D$ with the help of certain simplicial complexes, and study the linear codes $C_D$ over $I$. We obtain the Lee-weight distributions for these codes by using Boolean functions. By considering a Gray map, we obtain certain binary linear codes, and their weight distributions. Most of these binary codes turn out to be self-orthogonal. By choosing the defining set appropriately, we produce an infinite family of optimal codes. Moreover, most of the binary codes obtained here are minimal. We give a few examples to illustrate our results.\par
		The remaining sections of this article are arranged as follows. Preliminaries are presented in the next section. By using simplicial complexes, we construct linear $I$-codes and obtain their Lee-weight distributions in Section \ref{section3}. Section \ref{section4} studies Gray images of these linear $I$-codes. In this section, we obtain a class of distance optimal codes and establish certain conditions under which the Gray images of linear $I$-codes are minimal and self-orthogonal. In Section \ref{section5}, we conclude this article.

		\section{Definitions and Preliminaries}\label{section2}
		Suppose $I$ is the ring given by $I = \langle a, b ~\vert ~ 2a = 0, 2b = 0, a^2 = b,ab = 0\rangle $ (notation as in \cite{Fine}). Suppose the underlying set of $I$ is $\{0, a, b, c\}$. Then the addition and multiplication tables of $I$ are given by Table \ref{additionTable} and \ref{multiplicationTable} respectively.
		\begin{table}[h]
			\begin{center}			
			\begin{tabular}{c|c|c|c|c}
				$+$ & 0 & a & b & c \\
				\hline
				0 & 0 & a & b & c \\
				\hline
				a & a & 0 & c & b \\
				\hline
				b & b & c & 0 & a \\
				\hline
				c & c & b & a & 0 \\
			\end{tabular}
		\end{center}
	    \caption{Addition table of the ring $I$}
	    \label{additionTable}
		\end{table}
		\begin{table}[h]
			\begin{center}
			\begin{tabular}{c|c|c|c|c}
		 $\times$ & 0 & a & b & c \\
				\hline
				0 & 0 & 0 & 0 & 0 \\
				\hline
				a & 0 & b & 0 & b \\
				\hline
				b & 0 & 0 & 0 & 0 \\
				\hline
				c & 0 & b & 0 & b \\
			\end{tabular}
		    \end{center}
	      \caption{Multiplication table of the ring $I$}
	      \label{multiplicationTable}
		\end{table}
	    It is easy to observe from the multiplication table that $I$ is commutative and non-unital. Consider the following action of $\mathbb{F}_2$ on $I$: $x0 = 0x = 0$ and $x1 = 1x = x$ for all $x \in I$. Then every element of $I$ can be expressed as $as+bt$ for $s, t \in \mathbb{F}_2$ (see \cite{AlahmadiI1}).
		\begin{lemma}\cite{AlahmadiI1}
			For any $n \in \mathbb{N}$, $I^n = a\mathbb{F}_2^n + b\mathbb{F}_2^n$ and the sum is direct.	    	
		\end{lemma}		        
        Suppose $\Phi : I \longrightarrow \mathbb{F}_2^2$ is the \textit{Gray map} given by 
		\begin{equation}\label{PhiEq}
			\Phi(as + bt) = (t, s+t).
		\end{equation}
		This extends to a map from $I^m$ to $(\mathbb{F}_2^{m})^2$ component-wise for any $m \in \mathbb{N}$.
		\begin{definition}
			An $I$-submodule $C$ of $I^m$ is called a \textit{linear $I$-code} of length $m$.
		\end{definition}
		Let $v, w \in \mathbb{F}_2^m$. Then the \textit{Hamming weight} of $v$ denoted by $wt_{H}( v )$ is the number of non-zero entries in $v$. The \textit{Hamming distance} between $v$ and $w$ is $d_H(v, w) = wt_{H}(v-w)$.\\
		Let $x=a\alpha + b\beta$, $y=a\alpha' + b\beta' \in I^m$, where $\alpha, \beta, \alpha', \beta'\in \mathbb{F}_2^m$. Then the \textit{Lee weight} of $x$ is $wt_{Lee}(x) = wt_{H}(\Phi(x)) = wt_{H}(\beta) + wt_{H}(\alpha +\beta )$. The \textit{Lee distance} between $x$ and $y$ is $d_{Lee}(x, y) = wt_{Lee}(x-y)$ so that $\Phi$ is an isometry. Note that the image of a linear $I$-code under the above Gray map is a binary linear code. Suppose $C$ is a linear $I$-code of length $m$. Let $A_i$ be the cardinality of the set that contains all codewords of $C$ having Lee weight $i$, $0\leq i\leq 2m$. Then the homogeneous polynomial in two variables \[Lee_C(X, Y) = \sum_{c\in C}X^{2m-wt_{Lee}(c)}Y^{wt_{Lee}(c)}\] is called the \textit{Lee weight enumerator} of $C$ and the string $(1, A_1,\dots ,A_{2m})$ is called the \textit{Lee weight distribution} of $C$. In a similar way, we can define Hamming weight enumerator and Hamming weight distribution of a linear code over a finite field. In addition, if the total number of $i\geq 1$ such that $A_i\neq 0$ is $l$, then $C$ is called an $l$-\textit{weight linear code}. Every $1$-weight linear code is an equidistant linear code. Bonisoli characterized all equidistant linear codes over finite fields in \cite{Bonisoli}.
       \begin{theorem}\cite{Bonisoli}\textnormal{(Bonisoli)}\label{Bonisoli}
       	Suppose $C$ is a equidistant linear code over $\mathbb{F}_q$. Then $C$ is equivalent to the $r$-fold replication of a simplex code, possibly with added $0$-coordinates.
       \end{theorem}
       An $[n, k, d]$-linear code $C$ is called \textit{distance optimal} if there exist no $[n,k,d+1]$-linear codes (see \cite{wchuffman}). Next we recall the Griesmer bound.
       \begin{lemma}\label{griesmerbound}\cite{Griesmer}
       	\textnormal{(Griesmer Bound)} If $C$ is an $[n,k,d]$-linear code over $\mathbb{F}_q$, then we have
       	\begin{equation}\label{GriesmerBound}
       		\sum\limits_{i=0}^{k-1}\left\lceil \frac{d}{q^i}\right\rceil \leq n,
       	\end{equation}  	
       	where $\lceil \cdot \rceil$ denotes the ceiling function.
       \end{lemma}
       A linear code is called a \textit{Griesmer code} if equality holds in Equation \eqref{GriesmerBound}. Note that every Griesmer code is distance optimal, but not conversely.\\
      For $m\in \mathbb{N}$, we shall write $[m]$ to denote the set $\{1, 2,\dots ,m\}$ and $w\in \mathbb{F}_{2}^m$. Then the set $\textnormal{Supp}(w)=\{ i\in [m]: w_i = 1\}$ is called the \textit{support} of $w$. Note that the Hamming weight of $w\in \mathbb{F}_2^m$ is $wt_H(w)=\vert \textnormal{Supp}(w)\vert $. For $v,w\in \mathbb{F}_{2}^m$, one says that $v$ \textit{covers} $w$ if $\textnormal{Supp}(w)\subseteq \textnormal{Supp}(v)$. If $v$ covers $w$, we write $w\preceq v$.\\
	  Consider the map $\psi: \mathbb{F}_2^m\longrightarrow 2^{[m]}$ is defined as $\psi(w)=\textnormal{Supp}(w)$, where $2^{[m]}$ denotes the power set of $[m]$. Note that $\psi$ is a bijective map. Now onwards, we will write $w$ instead of Supp($w$) whenever we require.
	\begin{definition}
		Let $C$ be a linear code over $\mathbb{F}_2$. An element $v\in C\setminus \{0\}$ is called \textit{minimal} if $w\preceq v$ and $w\in C\setminus \{0\}$ $\implies$ $w = v$. If each nonzero codeword of $C$ is minimal then $C$ is called a \textit{minimal code}.
	\end{definition}
    Now we recall a result from \cite{suffConMinimalcode} which is a sufficient condition for a linear code over $\mathbb{F}_q$ to be minimal.
    \begin{lemma}\label{minimal_lemma}
    	\cite{suffConMinimalcode}\textnormal{(Ashikhmin-Barg)}
    	Let $C$ be a linear code over $\mathbb{F}_q$ with $wt_o$ and $wt_\infty$ as minimum and maximum weights of its non-zero codewords. If $\frac{wt_0}{wt_{\infty}}> \frac{q-1}{q}$, then $C$ is minimal.
    \end{lemma}
    Here we recall the definition of simplicial complex.
	\begin{definition}
		A subset $\Delta$ of $\mathbb{F}_{2}^m$ is called a \textit{simplicial complex} if $v\in \Delta, w\in \mathbb{F}_2^m$ and $w\preceq  v$ $\implies$ $w\in \Delta$. An element $v\in \Delta$ is called a \textit{maximal element of} $\Delta$ if for $w\in \Delta$, $v\preceq w$ $\implies$ $v=w$.
	\end{definition} 
    A simplicial complex can have more than one maximal elements. Let $M\subseteq [m]$. The simplicial complex generated by $M$ is denoted by $\Delta_{M}$ and is defined as 
	\begin{equation}
		\Delta_{M}=\{w\in \mathbb{F}_{2}^m \vert \textnormal{ Supp}(w)\subseteq M\}=\psi^{-1}(2^M).
	\end{equation}
	Note that $\psi^{-1}(M)$ is the only maximal element of $\Delta_M$, and $\vert \Delta_{M}\vert = 2^{\vert M\vert}$. Here $\Delta_{M}$ is a vector space over $\mathbb{F}_2$ of dimension $\vert M \vert$.\par
	Given a subset $P$ of $\mathbb{F}_{2}^m$, define the polynomial (referred to as an \textit{$m$-variable generating function}, see \cite{Chang}) $\mathcal{H}_{P}(y_1, y_2,\dots , y_m)$ by
	\begin{equation}
		\mathcal{H}_P(y_1,y_2,\dots ,y_m)=\sum\limits_{v\in P}\prod_{i=1}^{m}y_i^{v_i}\in \mathbb{Z}[y_1,y_2,\dots ,y_m],
	\end{equation}
	where $v=(v_1,\dots ,v_m)$ and $\mathbb{Z}$ denotes the ring of integers.\\	
	We recall a lemma from \cite{Chang}.
	\begin{lemma}\cite{Chang}\label{generatinglemma}
		Suppose $\Delta\subseteq \mathbb{F}_{2}^m$ is a simplicial complex and $\mathcal{F}$ consists of its maximal elements. Then 
		\begin{equation}
			\mathcal{H}_{\Delta}(y_1,y_2,\dots ,y_m)=\sum\limits_{\emptyset\neq S\subseteq\mathcal{F}}(-1)^{\vert S\vert +1}\prod_{i\in \cap S}(1+y_i),
		\end{equation}
		where $\cap S=\bigcap\limits_{F\in S}\textnormal{Supp}(F)$. In particular, we have
		\begin{equation*}
			\vert \Delta\vert =\sum\limits_{\emptyset\neq S\subseteq \mathcal{F}}(-1)^{\vert S\vert +1}2^{\vert \cap S\vert}.
		\end{equation*}
	\end{lemma}
	\begin{example}
		Consider the simplicial complex
		\begin{equation*}
			\Delta = \{(0, 0, 0, 0), (0, 0, 1, 0), (0, 0, 0, 1), (0, 0, 1, 1), (0, 1, 0, 0), (0, 1, 0, 1)\}.
		\end{equation*}
		Then $\mathcal{F} = \{F_1, F_2\}$ where $F_1 = (0, 0, 1, 1)$ and $F_2 = (0, 1, 0, 1)$. So 
		\begin{equation*}
			\begin{split}
				\mathcal{H}_{\Delta}(y_1, y_2, y_3, y_4) & = \prod_{i\in F_1}(1+y_i) + \prod_{i\in F_2}(1+y_i) - \prod_{i\in F_1\cap F_2}(1+y_i)\\
				& = 1 + y_2 + y_3 + y_4 + y_2y_4 + y_3y_4.				
			\end{split}
		\end{equation*}
	 and $\vert \Delta \vert = 6.$
	\end{example}
    For $M\subseteq [m]$, define a Boolean function $\Psi(\cdot \vert M): \mathbb{F}_2^m \longrightarrow \mathbb{F}_2$  by
    \begin{equation}\label{BooleanFunction}
    	\Psi(\alpha\vert M)=\prod_{i\in M}(1-\alpha_i)=\begin{cases}
    		1, \text{  if } \textnormal{ Supp}(\alpha)\cap M=\emptyset;\\
    		0, \text{  if } \textnormal{ Supp}(\alpha)\cap M\neq \emptyset.
    	\end{cases}
    \end{equation}
    We recall a lemma from \cite{Sagar_Sarma2}.
    \begin{lemma}\cite{Sagar_Sarma2}\label{countingLemma}
    	Suppose $M$ is a subset of $[m]$. Then
    	\begin{enumerate}
    			\item \label{1.1}
    			$\vert \{v\in \mathbb{F}_{2}^m : \Psi(v \vert M) = 1\}\vert = 2^{m-\vert M\vert}$.
    			\item \label{1.2}
    			$\vert \{v\in \mathbb{F}_{2}^m : \Psi(v \vert M) = 0\}\vert = (2^{\vert M\vert }-1)\times 2^{m-\vert M\vert}$.
    	\end{enumerate}
    \end{lemma}    
    
    Now we recall a result from \cite{shi_nonchain}.
    \begin{lemma}\cite{shi_nonchain}\label{relation_with_complement}
    	Let $\alpha \in \mathbb{F}_2^m$ and let $\Delta_{M}$ be the simplicial complex generated by $M \subseteq [m]$. Then
    	\begin{equation*}
    		\begin{split}
    			\sum_{t\in \Delta_{M}^{\textnormal{c}}}(-1)^{\alpha t}=2^{m}\delta_{0, \alpha}-\sum_{t\in \Delta_{M}}(-1)^{\alpha t},
    		\end{split}
    	\end{equation*}
    	where $\delta$ is the Kronecker delta function, and $\Delta_{M}^{\textnormal{c}}=\mathbb{F}_2^m\setminus \Delta_{M}$, the complement of $\Delta_{M}$.
    \end{lemma}
	In the forthcoming sections, we study the algebraic structures of linear $I$-codes and their binary Gray images.
	\section{Construction of linear $I$-codes via simplicial complexes}\label{section3}
	  For any $n, m\in \mathbb{N}$, let $D=\{d_1<d_2<\dots <d_n\}\subseteq I^m$ be an ordered multiset.
	  Let $D_i\subseteq \mathbb{F}_2^m, i=1, 2$. Assume that $D=aD_1 + bD_2\subseteq I^m$. Define
		\begin{equation}\label{C_Ddefinitionleft}
			C_D=\{c_D(v)=\big(v\cdot d\big)_{d\in D} ~\vert~ v \in I^m\}.
		\end{equation}
	   where $x\cdot y=\sum\limits_{i=1}^{m}x_iy_i$ for $x,y\in I^m$.\\
	   Then $C_D$ is a linear $I$-code of length $\vert D\vert $. The ordered set $D$ is called the \textit{defining set} of $C_{D}$. Throughout this article we consider a defining set to be an ordered multiset. Note that on changing the ordering of $D$ we will get a code which is permutation equivalent (see \cite{wchuffman}) to $C_D$. Observe that $c_{D}: I^m\longrightarrow C_{D}\subseteq I^{\vert D \vert}$ defined by
	   \begin{equation}\label{c_DMap}
	   	 c_{D}(v)=\big(v\cdot d\big)_{d\in D}
	   \end{equation}
      is a surjective homomorphism of $I$-modules.\\	   
	   Assume that $x=a\alpha + b\beta \in I^m$ and $d=at_1 + bt_2\in D = aD_1 + bD_2$, where $\alpha, \beta \in \mathbb{F}_2^m$ and $t_i\in D_i, i=1, 2$. Then the Lee weight of $c_{D}(x)$ is
	   \begin{equation*}
	   	\begin{split}
	   		wt_{Lee}(c_{D}(x)) = & wt_{Lee}\big(\big(\big(a\alpha + b\beta \big)\cdot \big(at_1 + bt_2 \big)\big)_{t_i\in D_i}\big)\\
	   		                   = & wt_{Lee}\big(\big(b(\alpha t_1)\big)_{t_i\in D_i}\big)\\
	   		                   = & wt_{H}\big(\Phi\big((b(\alpha t_1))_{t_i\in D_i}\big)\big)\\
	   		                   = & wt_{H}\big((\alpha t_1)_{t_i\in D_i}\big) + wt_{H}\big((\alpha t_1)_{t_i\in D_i}\big)\\
	   		                   = & 2wt_{H}\big((\alpha t_1)_{t_i\in D_i}\big).
	   	\end{split}
	   \end{equation*}   
       Now if $v \in \mathbb{F}_{2}^m$, then $wt_H(v)=0 \iff v=\textbf{0}\in \mathbb{F}_{2}^m $. Hence,
       \begin{equation}\label{keyeq1}
   	     \begin{split}
   		   wt_{Lee}(c_{D}(x))& =  2\big[\vert D \vert -\frac{1}{2}\sum\limits_{t_1\in D_1}\sum\limits_{t_2\in D_2}\big(1+(-1)^{\alpha t_1 } \big)\big]\\
   		                     & =  \vert D\vert - \sum\limits_{t_1\in D_1}(-1)^{\alpha t_1}\sum\limits_{t_2\in D_2}(1).
   	     \end{split}
       \end{equation}
       For $P \subseteq \mathbb{F}_2^m$ and $\alpha \in \mathbb{F}_2^m$, we define
       \begin{equation}\label{chi_def}
       	 \chi_{\alpha}(P) = \sum\limits_{t\in P}(-1)^{\alpha t}.
       \end{equation}
       Note that, for $\alpha \in \mathbb{F}_2^m$ and $\emptyset \neq M \subseteq [m]$, we have
       \begin{equation}\label{keyeq2}
       	   \begin{split}
       	   	  \chi_{\alpha}(\Delta_{M})&=\sum\limits_{t\in \Delta_{M}}(-1)^{\alpha t}\\
       	   	  &= \mathcal{H}_{\Delta_{M}}\big((-1)^{\alpha_1}, (-1)^{\alpha_2},\dots , (-1)^{\alpha_m}\big)\\
       	   	  &=\prod_{i\in M}(1+(-1)^{\alpha_i})=\prod_{i\in M}(2-2\alpha_i) \textnormal{ (By Lemma }\ref{generatinglemma})\\
       	   	  &=2^{\vert M \vert }\prod_{i\in M}(1-\alpha_i) = 2^{\vert M \vert}\Psi(\alpha\vert M),
       	   \end{split}
       \end{equation}
      where $\Psi(\cdot \vert M)$ is the Boolean function defined by Equation (\ref{BooleanFunction}).\\
      The following result describes Lee weight distributions of $C_D$ for various choices of $D$.
		\begin{theorem}\label{main}
			Suppose that $m\in \mathbb{N}$ and $M, N\subseteq [m]$.
			\begin{enumerate}
				\item \label{proposition:1}
				Let $D = a\Delta_M + b\Delta_N\subseteq I^m$. Then $C_D$ is a linear $I$-code of length $\vert D\vert = 2^{\vert M\vert + \vert N \vert }$ and size $2^{\vert M \vert}$. The Lee weight distribution of $C_D$ is displayed in Table \ref{table:1}.
					\begin{table}[]
						\centering
							\begin{tabular}{  c | c  }
								\hline
								Lee weight  & Frequency \\
								\hline
								$2^{\vert M \vert + \vert N \vert }$ & $2^{2m-\vert M\vert}\times (2^{\vert M \vert}-1)$\\
								\hline
								$0$  & $2^{2m-\vert M\vert}$\\
								\hline
							\end{tabular}
						\caption{Lee weight distribution in Theorem \ref{main} (\ref{proposition:1})}
						\label{table:1}	
					\end{table}

				\item \label{proposition:2}
				Let $D = a\Delta^{\textnormal{c}}_M + b\Delta_N \subseteq I^m$. Then $C_D$ is a linear $I$-code of length $\vert D\vert = (2^m-2^{\vert M\vert})\times 2^{ \vert N \vert}$ and size $2^{m}$. The Lee weight distribution of $C_D$ is displayed in Table \ref{table:2}.
					\begin{table}[]
						\centering
							\begin{tabular}{  c | c  }
								\hline
								Lee weight  & Frequency \\
								\hline								
								$2^{m + \vert N\vert}$     & $2^{m}\times (2^{m-\vert M\vert}- 1)$\\
								\hline
								$(2^m-2^{\vert M \vert})\times 2^{\vert N \vert}$ & $2^{2m-\vert M \vert}\times (2^{\vert M \vert} -1)$\\
								\hline
								$0$ & $2^m$\\
								\hline
							\end{tabular}
						\caption{Lee weight distribution in Theorem \ref{main} (\ref{proposition:2})}
						\label{table:2}		
					\end{table}
				
				\item \label{proposition:3}
				Let $D = a\Delta_M + b\Delta^{\textnormal{c}}_N \subseteq I^m$. Then $C_D$ is a linear $I$-code of length $\vert D\vert = 2^{\vert M \vert}\times(2^m-2^{\vert N \vert})$ and size $2^{\vert M\vert }$. The Lee weight distribution of $C_D$ is displayed in Table \ref{table:3}.
				\begin{table}[]
					\centering
						\begin{tabular}{  c | c  }
							\hline
							Lee weight  & Frequency \\
							\hline
							$(2^m - 2^{\vert N \vert})\times 2^{\vert M \vert}$ & $2^{2m-\vert M\vert }\times (2^{\vert M \vert} -1)$\\
							\hline
							$0$ & $ 2^{2m - \vert M \vert }$\\
							\hline
						\end{tabular}
					\caption{Lee weight distribution in Theorem \ref{main} (\ref{proposition:3})}
					\label{table:3}		
				\end{table}
				
				\item \label{proposition:4}
				Let $D = a\Delta^{\textnormal{c}}_M + b\Delta_N^{\textnormal{c}}\subseteq I^m$. Then $C_D$ is a linear $I$-code of length $\vert D\vert = (2^m-2^{\vert M\vert })(2^m-2^{\vert N\vert })$ and size $2^{m}$. The Lee weight distribution of $C_D$ is displayed in Table \ref{table:4}.
				\begin{table}[]
					\centering
						\begin{tabular}{  c | c  }
							\hline
							Lee weight    & Frequency \\
							\hline
							$2^m\times (2^m - 2^{\vert N \vert})$ & $2^{m}\times (2^{m-\vert M\vert} - 1)$\\
							\hline
							$(2^{m} - 2^{\vert M \vert})(2^{m}- 2^{\vert N \vert})$ & $2^{2m-\vert M \vert}\times (2^{\vert M \vert} -1)$\\
							\hline
							$0$ & $2^m$\\
							\hline
						\end{tabular}
					\caption{Lee weight distribution in Theorem \ref{main} (\ref{proposition:4})}
					\label{table:4}
				\end{table}	
			 \item \label{proposition:5}
			 Let $D = a\Delta_{M} + b\Delta_N$ so that $D^{\textnormal{c}} =\big(a\Delta_{M}^{\textnormal{c}} + b\mathbb{F}_2^m\big)\bigsqcup \big( a\Delta_{M} + b\Delta_N^{\textnormal{c}}\big)$, where $\bigsqcup$ denotes disjoint union. Then $C_{D^{\textnormal{c}}}$ is a linear $I$-code of length $\vert D^{\textnormal{c}} \vert =2^{2m} - 2^{\vert M \vert + \vert N \vert}$ and size $2^{m}$. The Lee weight distribution of $C_{D^{\textnormal{c}}}$ is displayed in Table \ref{table:5}.
			 \begin{table}[h]
			 	\centering
			 		\begin{tabular}{  c | c  }
			 			\hline
			 			Lee weight    & Frequency \\
			 			\hline
			 			$2^{2m}$ & $2^m\times (2^{m-\vert M \vert} -1)$\\
			 			\hline
			 			$2^{2m} - 2^{\vert M \vert + \vert N \vert }$ & $2^{2m - \vert M \vert}\times (2^{\vert M \vert} -1)$\\
			 			\hline
			 			$0$ & $2^m$\\
			 			\hline
			 		\end{tabular}
			 	\caption{Lee weight distribution in Theorem \ref{main} (\ref{proposition:5})}
			 	\label{table:5}
			 \end{table}
			\end{enumerate}
		\end{theorem}
         \proof We discuss the proof of part (\ref{proposition:4}). The other parts can be proved in a similar way.\\
         Observe the length of $C_D$ is $\vert D\vert = (2^m -2^{\vert M \vert})(2^m - 2^{\vert N \vert})$. Let $x = a\alpha + b\beta \in I^m$. By Equation (\ref{keyeq1}) and (\ref{keyeq2}), we have
         \begin{equation}
         	\begin{split}
         		wt_{Lee}(c_{D}(x)) & = (2^m - 2^{\vert M \vert})(2^m - 2^{\vert N \vert}) - \big[2^m\delta_{0, \alpha} - 2^{\vert M \vert}\Psi(\alpha \vert M)\big]\times (2^m - 2^{\vert N \vert}).
         	\end{split}
         \end{equation}
        Here we discuss the following cases.
        \begin{enumerate}
        	\item If $\alpha = 0$, then $wt_{Lee}(c_D(x)) = 0.$\\
            In this case, $\#\alpha =1, \#\beta = 2^m$.\\
            Therefore, $\# x = 2^m$.
            \item If $\alpha \neq 0$, then $wt_{Lee}(c_D(x)) =(2^m - 2^{\vert M \vert})(2^m - 2^{\vert N \vert}) + 2^{\vert M \vert}(2^m - 2^{\vert N \vert})\Psi(\alpha \vert M)$.            
            \begin{enumerate}
            	\item If $\Psi(\alpha \vert M) = 0 $ then $wt_{Lee}(c_D(x)) = (2^m - 2^{\vert M \vert})(2^m - 2^{\vert N \vert}).$\\
            	In this case, by using Lemma \ref{countingLemma},  we get\\
            	 $\#\alpha = (2^{\vert M \vert} -1)\times 2^{m-\vert M \vert}$,\\
            	$\#\beta = 2^m$.\\
            	 Therefore, $\#x = 2^{2m-\vert M \vert}(2^{\vert M \vert } -1)$.
            	
            	\item If $\Psi(\alpha \vert M) = 1$ then $wt_{Lee}(c_D(x)) =2^m(2^m - 2^{\vert N \vert}).$\\ 
            	In this case, by using Lemma \ref{countingLemma},  we get \\
            	$\#\alpha =(2^{m - \vert M \vert }-1)$,\\
            	$\#\beta = 2^m$.\\
            	Therefore, $\#x= 2^m(2^{m - \vert M \vert }-1)$.  
            \end{enumerate}
        \end{enumerate}
         Based on the above computations, we obtain Table \ref{table:4}.\\
         By using Table \ref{table:4}, we have $\vert \ker(c_D)\vert = \vert \{v\in I^m : v\cdot d = 0 ~\forall ~d \in D\}\vert = 2^m$. By the first isomorphism theorem, we have $\vert C_D \vert = \frac{\vert I^m \vert}{\vert \ker(c_D)\vert} = 2^m$. \qed
        \section{Gray images of linear $I$-codes}\label{section4}
        In this section, we study the Gray images of linear $I$-codes which are discussed in Section \ref{section3}.\\
        Now we recall a result (see Theorem $1.4.8 ~(ii)$ of \cite{wchuffman}) for self-orthogonality of binary linear codes.
        \begin{theorem}\cite{wchuffman}\label{orthogonal_lemma}
        	If the Hamming weight of every non-zero element of a binary linear code $C$ is a multiple of $4$, then $C$ is self-orthogonal.
        \end{theorem}
        By using the above result, we give a sufficient condition for the binary Gray images $\Phi(C_D)$ to be self-orthogonal for each $C_D$ discussed in Theorem \ref{main}.
        \begin{proposition}\label{mainright_Ortho}
        	Assume that $C_D$ is as in Theorem \ref{main}. Then its Gray image $\Phi(C_D)$ is a binary self-orthogonal code provided $\vert M \vert + \vert N \vert \geq 2$.
        \end{proposition}
        By using Lemma \ref{minimal_lemma} and Lemma \ref{griesmerbound} respectively, we find minimal and optimal linear codes among the Gray images of the codes studied in Theorem \ref{main}.
        \begin{theorem}\label{GrayImageC_D}
        	Let $m\in \mathbb{N}$ and $M, N\subseteq [m]$. Suppose $\Phi$ is the map defined by Equation (\ref{PhiEq}).
        	\begin{enumerate}
        		\item \label{item_Gray1}
        		Let $D = a\Delta_{M} + b\Delta_{N}\subseteq I^m$. Then $\Phi(C_D)$ is a binary $[2^{\vert M \vert + \vert N \vert+1 }, \vert M \vert, 2^{\vert M \vert + \vert N \vert}]$-linear $1$-weight code.
        		\item \label{item_Gray2}
        		Let $D = a\Delta^{\textnormal{c}}_{M} + b\Delta_{N}\subseteq I^m$. Then $\Phi(C_D)$ is a binary $[(2^m-2^{\vert M \vert })2^{\vert N \vert +1}, m, (2^m-2^{\vert M \vert})2^{\vert N \vert}]$-linear $2$-weight code. It is minimal provided $\vert M \vert \leq m-2$.
        		\begin{enumerate}
        			\item \label{item_Gray2.1}
        			Let $1 \leq \vert M \vert + \vert N \vert \leq m-1$ and $\theta_1 = 2^{\vert N \vert +1}-1$. If $0< \theta_1 < \vert M \vert + \vert N \vert +1$ then $\Phi(C_D)$ is optimal with respect to the Griesmer bound.
        			\item \label{item_Gray2.2}
        			Let $m \leq \vert M \vert + \vert N \vert \leq 2m-1$ and $\theta_2 = 2^{\vert M \vert + \vert N \vert +1 -m}(2^{m-\vert M \vert} -1)$. If $0 < \theta_2 < m$ then $\Phi(C_D)$ is optimal with respect to the Griesmer bound.
        		\end{enumerate}
        		\item \label{item_Gray3}
        		Let $D = a\Delta_{M} + b\Delta^{\textnormal{c}}_{N}\subseteq I^m$. Then $\Phi(C_D)$ is a binary $[2^{\vert M \vert + 1 }(2^m - 2^{\vert N \vert}), \vert M \vert, 2^{\vert M \vert }(2^m - 2^{\vert N \vert})]$-linear $1$-weight code.
        		
        		\item \label{item_Gray4}
        		Let $D = a\Delta^{\textnormal{c}}_{M} + b\Delta^{\textnormal{c}}_{N}\subseteq I^m$. Then $\Phi(C_D)$ is a binary $[2(2^m-2^{\vert M \vert})(2^m - 2^{\vert N \vert}), m, (2^m-2^{\vert M \vert})(2^m - 2^{\vert N \vert})]$-linear $2$-weight code. If $\vert M \vert \leq m-2$ then $C_D$ is a minimal code.
        		\item \label{item_Gray5}
        		Let $D = a\Delta_{M} + b\Delta_{N}\subseteq I^m$ so that $D^{\textnormal{c}} = (a\Delta^{\textnormal{c}}_{M} + b\mathbb{F}_2^m)\bigsqcup (a\Delta_{M} + b\Delta^{\textnormal{c}}_{N})$. Then $\Phi(C_{D^{\textnormal{c}}})$ is a binary $[2(2^{2m} - 2^{\vert M \vert +\vert N \vert}), m, (2^{2m} - 2^{\vert M \vert +\vert N \vert})]$-linear $2$-weight code. If $\vert M \vert + \vert N \vert \leq 2m-2$ then $\Phi(C_{D^{\textnormal{c}}})$ is a minimal code.
        	\end{enumerate}
        \end{theorem}
        \proof We discuss case (\ref{item_Gray2}) and omit others.\\
        From Table (\ref{table:2}), we have $wt_0 = (2^m - 2^{\vert M\vert })\times 2^{\vert N \vert}$ and $wt_ {\infty} = 2^{m + \vert N\vert}$.\\
        Now, $\frac{w_0}{w_{\infty}} = \frac{(2^m - 2^{\vert M\vert })\times 2^{\vert N \vert}}{2^{m + \vert N\vert}} > \frac{1}{2} \iff \vert M \vert \leq m-2$. Therefore, by Lemma \ref{minimal_lemma}, $\Phi(C_D)$ is minimal if $\vert M \vert \leq m-2$.\\
        (\ref{item_Gray2.1}) Suppose $1 \leq \vert M \vert + \vert N \vert \leq m-1$ and $\theta_1 = 2^{\vert N \vert +1} -1$. The parameters of the code $\Phi(C_D)$ are given by $n=(2^m-2^{\vert M \vert})2^{\vert N \vert +1}, k=m, d= (2^m-2^{\vert M \vert})2^{\vert N \vert}$.\\
        Note that
        \begin{equation*}
        	\begin{split}
        		\sum_{i=0}^{k-1}\left\lceil \frac{d}{2^i}\right\rceil & =	\sum_{i=0}^{m-1}\left\lceil \frac{2^{m+\vert N \vert}-2^{\vert M \vert + \vert N \vert}}{2^i}\right\rceil \\
        		& = \sum_{i=0}^{m-1}\frac{2^{m+\vert N \vert}}{2^i}-	\sum_{i=0}^{m-1}\left\lfloor \frac{2^{\vert M \vert + \vert N \vert}}{2^i}\right\rfloor\\
        		& = (2^{m+\vert N \vert +1} - 2^{\vert N \vert +1})-(2^{\vert M \vert + \vert N \vert +1} -1)\\
        		&= (2^{m+\vert N \vert +1} - 2^{\vert M \vert + \vert N \vert +1})-( 2^{\vert N \vert +1} -1)\\
        		& = n-\theta_1.
        	\end{split}
        \end{equation*}
        By Griesmer bound, $\sum\limits_{i=0}^{k-1}\lceil \frac{d}{2^i}\rceil \leq n \iff \theta_1 > 0$.\\
        On the other hand,
        \begin{equation*}
        	\begin{split}
        		\sum_{i=0}^{m-1}\left\lceil \frac{2^{m+\vert N \vert}-2^{\vert M \vert + \vert N \vert} +1}{2^i}\right\rceil 
        		& = \sum_{i=0}^{m-1}\frac{2^{m+\vert N \vert}}{2^i}-\sum_{i=0}^{\vert M \vert + \vert N \vert}\left\lfloor \frac{2^{\vert M \vert + \vert N \vert}}{2^i} -\frac{1}{2^i}\right\rfloor \\
        		& -\sum_{i=\vert M \vert + \vert N \vert+1}^{m-1}\left\lfloor \frac{2^{\vert M \vert + \vert N \vert}}{2^i} -\frac{1}{2^i}\right\rfloor\\
        		& = (2^{m+\vert N \vert +1} - 2^{\vert N \vert +1})-\big[(2^{\vert M \vert + \vert N \vert +1} -1)\\ &~ - (\vert M \vert + \vert N \vert+1)\big]-0\\
        		& = (2^{m+\vert N \vert +1} - 2^{\vert M \vert + \vert N \vert +1})-(2^{\vert N \vert +1}-1)+\vert M \vert + \vert N \vert+1\\
        		& = n-\theta_1 + \vert M \vert + \vert N \vert+1.
        	\end{split}
        \end{equation*}
        Therefore $\sum\limits_{i=0}^{k-1}\lceil \frac{d + 1}{2^i}\rceil > n \iff \theta_1 < \vert M \vert + \vert N \vert+1$. This completes the proof of (\ref{item_Gray2.1}).\\
        (\ref{item_Gray2.2}) Suppose $m \leq \vert M \vert + \vert N \vert \leq 2m-1$ and $\theta_2 = 2^{\vert M \vert + \vert N \vert - m +1}(2^{m-\vert M \vert}-1)$. Now we have
        \begin{equation*}
        	\begin{split}
        		\sum_{i=0}^{k-1}\left\lceil \frac{d}{2^i}\right\rceil & =	\sum_{i=0}^{m-1}\left\lceil \frac{2^{m+\vert N \vert}-2^{\vert M \vert + \vert N \vert}}{2^i}\right\rceil \\
        		& = \sum_{i=0}^{m-1}\frac{2^{m+\vert N \vert}}{2^i}-	\sum_{i=0}^{m-1}\left\lfloor \frac{2^{\vert M \vert + \vert N \vert}}{2^i}\right\rfloor\\
        		& = (2^{m+\vert N \vert +1} - 2^{\vert N \vert +1})-(2^{\vert M \vert + \vert N \vert +1} - 2^{\vert M \vert + \vert N \vert -m +1})\\
        		& =  (2^{m+\vert N \vert +1} - 2^{\vert M \vert + \vert N \vert +1})-(2^{\vert N \vert +1} - 2^{\vert M \vert + \vert N \vert -m +1})\\
        		& = n-\theta_2.
        	\end{split}
        \end{equation*}
        By Griesmer bound, $\sum\limits_{i=0}^{k-1}\lceil \frac{d}{2^i}\rceil \leq n \iff \theta_2 > 0$.\\
        On the other hand,
        \begin{equation*}
        	\begin{split}
        		\sum_{i=0}^{m-1}\left\lceil \frac{2^{m+\vert N \vert}-2^{\vert M \vert + \vert N \vert} +1}{2^i}\right\rceil 
        		& = \sum_{i=0}^{m-1}\frac{2^{m+\vert N \vert}}{2^i}-\sum_{i=0}^{m-1}\left\lfloor \frac{2^{\vert M \vert + \vert N \vert}}{2^i} -\frac{1}{2^i}\right\rfloor \\        		
        		& = (2^{m+\vert N \vert +1} - 2^{\vert N \vert +1})-\big[(2^{\vert M \vert + \vert N \vert +1} - 2^{\vert M \vert + \vert N \vert -m +1}) -m\big]\\
        		& = (2^{m+\vert N \vert +1} - 2^{\vert M \vert + \vert N \vert +1})-(2^{\vert N \vert +1}- 2^{\vert M \vert + \vert N \vert -m +1})+m\\
        		& = n-\theta_2 + m.
        	\end{split}
        \end{equation*}
        Therefore $\sum\limits_{i=0}^{k-1}\lceil \frac{d + 1}{2^i}\rceil > n \iff \theta_2 < m$. This completes the proof of (\ref{item_Gray2.2}).\qed
        The following examples illustrate Theorem \ref{main}, Proposition \ref{mainright_Ortho} and Theorem \ref{GrayImageC_D}.        
        \begin{example}
        	\begin{enumerate}
        		\item Let $m = 6$, $M = \{2, 3\}, N=\{4, 5\}$. If $D = a\Delta_{M} + b\Delta_{N}$, then by using Theorem \ref{main}(\ref{proposition:1}) $C_D$ is a linear $I$-code with Lee weight enumerator $X^{32} + 3X^{16}Y^{16}$. By Proposition \ref{mainright_Ortho} and Theorem \ref{GrayImageC_D}(\ref{item_Gray1}), $\Phi(C_D)$ is a $[32, 2, 16]$ binary minimal and self-orthogonal code.
        		
        		\item 
        		\begin{enumerate}
        			\item Let $m = 5$, $M = \{1, 2, 3\}, N=\{4\}$. If $D = a\Delta_{M}^{\textnormal{c}} + b\Delta_{N}$, then by using Theorem \ref{main}(\ref{proposition:2}) $C_D$ is a linear $I$-code with Lee weight enumerator $X^{96} + 28 X^{48}Y^{48} + 3 X^{32} Y^{64}$. By Proposition \ref{mainright_Ortho} and Theorem \ref{GrayImageC_D}(\ref{item_Gray2.1}), $\Phi(C_D)$ is a $[96, 5, 48]$ binary minimal, self-orthogonal and distance optimal code. Note that optimality of $\Phi(C_D)$ can be verified using the Database \cite{BoundTable}.
        			
        			\item Let $m = 9$, $M = \{1, 2, 3, 4, 7, 8, 9\}, N=\{5, 6\}$. If $D = a\Delta_{M}^{\textnormal{c}} + b\Delta_{N}$, then by using Theorem \ref{main}(\ref{proposition:2}) $C_D$ is a linear $I$-code with Lee weight enumerator $X^{3072} + 508 X^{1536}Y^{1536} + 3 X^{1024}Y^{2048}$. By Proposition \ref{mainright_Ortho} and Theorem \ref{GrayImageC_D}(\ref{item_Gray2.2}), $\Phi(C_D)$ is a $[3072, 9, 1536]$ binary minimal, self-orthogonal and distance optimal code. Note that one can directly verify that there is no $[3072, 9, 1537]$-linear code over $\mathbb{F}_2$.
        		\end{enumerate}
        		
        		\item Let $m = 3$, $M = \{1, 2, 3\}, N=\{1, 2\}$. If $D = a\Delta_{M} + b\Delta_{N}^{\textnormal{c}}$, then by using Theorem \ref{main}(\ref{proposition:3}) $C_D$ is a linear $I$-code with Lee weight enumerator $X^{64} + 7 X^{32}Y^{32}$. By Proposition \ref{mainright_Ortho} and Theorem \ref{GrayImageC_D}(\ref{item_Gray3}), $\Phi(C_D)$ is a $[64, 3, 32]$ binary minimal and self-orthogonal code.

        		\item Let $m = 5$, $M = \{2, 3, 4\}, N=\{1, 2, 4, 5\}$. If $D = a\Delta_{M}^{\textnormal{c}} + b\Delta_{N}^{\textnormal{c}}$, then by using Theorem \ref{main}(\ref{proposition:4}) $C_D$ is a linear $I$-code with Lee weight enumerator $X^{768} + 28 X^{384}Y^{384} + 3 X^{256} Y^{512}$. By Proposition \ref{mainright_Ortho} and Theorem \ref{GrayImageC_D}(\ref{item_Gray4}), $\Phi(C_D)$ is a $[768, 5, 384]$ binary minimal and self-orthogonal code.
        		
        		\item Let $m = 4$, $M = \{2, 3, 4\}, N=\{1, 2, 4\}$. If $D= a\Delta_{M} + b\Delta_{N}$, then by using Theorem \ref{main}(\ref{proposition:5}) $C_{D^{\textnormal{c}}}$ is a linear $I$-code with Lee weight enumerator $X^{384} + 14 X^{192}Y^{192} + X^{128} Y^{256}$. By Proposition \ref{mainright_Ortho} and Theorem \ref{GrayImageC_D}(\ref{item_Gray5}), $\Phi(C_{D^{\textnormal{c}}})$ is a $[384, 4, 192]$ binary minimal and self-orthogonal code.
        	\end{enumerate}
        \end{example}
        \begin{remark}
        	By Theorem \ref{Bonisoli} all the binary $1$-weight codes of this article are simplex codes and hence they are distance optimal.
        \end{remark}
        \begin{table}[h]
        	\centering
        	\begin{adjustbox}{width=\textwidth}
        		
        		\begin{tabular}{c|c|c|c|c}
        			\hline
        			Result & $[n,k,d]$-code & $\#$Weight & Distance optimal & Minimal  \\        			
        			\hline
        			Theorem \ref{GrayImageC_D}(\ref{item_Gray1}) 	& $[2^{\vert M \vert + \vert N \vert +1}, \vert M \vert, 2^{\vert M \vert + \vert N \vert}]$ & 1 & Yes & Yes \\
        			\hline
        			Theorem \ref{GrayImageC_D}(\ref{item_Gray2.1}) & \multirow{2}{*}{$[(2^m - 2^{\vert M \vert})2^{\vert N \vert +1}, m, (2^m - 2^{\vert M \vert})2^{\vert N \vert}]$} & \multirow{2}{*}{2} & Yes, if $1\leq \theta_1 < \vert M \vert + \vert N \vert +1 \leq m$ & \multirow{2}{*}{Yes, if $\vert M \vert \leq m-2$} \\
        			\cline{1-1}\cline{4-4}
        			Theorem \ref{GrayImageC_D}(\ref{item_Gray2.2})  &  &  & Yes, $0 < \theta_2 < m \leq \vert M \vert + \vert N \vert$ & \\ 
        			\hline
        			Theorem \ref{GrayImageC_D}(\ref{item_Gray3}) & $[2^{\vert M \vert +1}(2^m - 2^{\vert N \vert}), \vert M \vert, 2^{\vert M \vert}(2^m - 2^{\vert N \vert})]$ & 1 & Yes & Yes \\
        			\hline
        			Theorem \ref{GrayImageC_D}(\ref{item_Gray4}) & $[2(2^m-2^{\vert M \vert})(2^m - 2^{\vert N \vert}), m, (2^m-2^{\vert M \vert})(2^m - 2^{\vert N \vert})]$ & 2 & $-$ & Yes, if $\vert M \vert \leq m-2 $ \\
        			\hline
        			Theorem \ref{GrayImageC_D}(\ref{item_Gray5}) 	& $[2(2^{2m} - 2^{\vert M \vert +\vert N \vert}), m, (2^{2m} - 2^{\vert M \vert +\vert N \vert})]$ & 2 & $-$ & Yes, if $\vert M \vert + \vert N \vert \leq 2m-2$ \\
        			\hline
        		\end{tabular}
        		
        	\end{adjustbox}
        	\caption{Binary linear codes from simplicial complexes obtained this paper}
        	\label{table:article}
        	
        \end{table}
        
		\section{Conclusion}\label{section5}
		In this article, we study few-weight linear $I$-codes via simplicial complexes having one maximal element. A Gray map is utilized to study the corresponding binary linear codes. We obtain their weight distributions by using Boolean functions. As a consequence, we achieve a class of distance optimal codes with respect to the Griesmer bound. Besides, we obtain many classes of binary minimal codes. Every family of the binary codes obtained here are self-orthogonal under certain mild conditions. For the reader's convenience, codes obtained in this article that have good parameters, are listed in in Table \ref{table:article}.
						


\begin{thebibliography}{10}
			\bibitem{AlahmadiE2} Alahmadi, A., Alkathiry, A., Altassan, A., Bonnecaze, A., Shoaib, H., Sol\'e, P.: The build-up construction of quasi self-dual codes over a non-unital ring. J. Algebra Appl. (2020), \url{https://doi.org/10.1142/S0219498822501432}
			
			\bibitem{AlahmadiI2} Alahmadi, A., Alkathiry, A., Altassan, A., Bonnecaze, A., Shoaib, H., Sol\'e, P.: The build-up construction over a commutative non-unital ring. Des. Codes Cryptogr. \textbf{90}(12), 3003-3010, (2022)
			
			\bibitem{AlahmadiI1} Alahmadi, A., Altassan, A., Basaffar, W., Bonnecaze, A., Shoaib, H., Sol\'e, P.: Quasi Type IV codes over a non-unital ring. Appl. Algebra Eng. Comm. Comput. \textbf{32}, 217-228, (2021) 
			
			
			\bibitem{AlahmadiE1} Alahmadi, A., Altassan, A., Basaffar, W., Shoaib, H., Bonnecaze, A., Sol\'e, P.: Type IV codes over a non-unital ring. J. of Algebra Appl. \textbf{21}(07), 2250142, (2022)
			
			\bibitem{suffConMinimalcode} Ashikhmin, A., Barg, A.: Minimal vectors in linear codes. IEEE Trans. Inf. Theory \textbf{44}(5), 2010-2017, (1998)					
			
			\bibitem{Ashikhmin_QECC} Ashikhmin, A., Knill, E.: Nonbinary quantum stabilizer codes. IEEE Trans. Inf. Theory \textbf{47}(7), 3065-3072, (2001)
			
			\bibitem{Bonisoli} Bonisoli, A.: Every equidistant linear code is a sequence of dual Hamming codes, Ars Combinatoria, \textbf{18}, 181–186, (1984)
			
			\bibitem{Few1} Calderbank, R., Kantor, W.M.: The geometry of two‐weight codes. Bulletin of the London Mathematical Society, \textbf{18}(2), 97-122, (1986)
			
			\bibitem{Calder_QECC} Calderbank, A.R., Rains, E.M., Shor, P.W., Sloane, N.J.A.: Quantum error correction via codes over GF (4). IEEE Transactions on Information Theory, \textbf{44}(4), 1369-1387, (1998)
			
			\bibitem{Chang} Chang, S., Hyun, J.Y.: Linear codes from simplicial complexes. Des. Codes Cryptogr. \textbf{86}(10), 2167-2181, (2018)
			
			
			\bibitem{Few2} Delsarte, P.: Weights of linear codes and strongly regular normed spaces. Discrete Mathematics, \textbf{3}(1-3), 47-64, (1972)
			
			\bibitem{Ding} Ding, C., Helleseth, T., Kl\o ve, T., Wang, X.: A generic construction of Cartesian authentication codes. IEEE Trans. Inf. Theory \textbf{53}(6), 2229-2235, (2007)
			
			
			\bibitem{Ding_Ding} Ding, K., Ding, C.: A class of two-weight and three-weight codes and their applications in secret sharing, IEEE Trans. Inf. Theory, \textbf{61}(11), 5835-5842, (2015)
			
			
			\bibitem{F2xF2} Dougherty, S.T., Gaborit, P., Harada, M., Munemasa, A., Sol\'e, P.: Type IV self-dual codes over rings, IEEE Trans. Inf. Theory \textbf{45}(7), 2345–2360, (1999)
			
			
			
			\bibitem{F2uF2} Dougherty, S.T., Gaborit, P., Harada, M., Sol\'e, P.: Type II Codes Over $\mathbb{F}_2 + u\mathbb{F}_2$, IEEE Trans. Inf. Theory \textbf{45}(1), 32–45, (1999)
			
			\bibitem{Du_Sihem} Du, Z., Li, C., Mesnager, S.: Constructions of self-orthogonal codes from hulls of BCH codes and their parameters. IEEE Trans. Inf. Theory, \textbf{66}(11), 6774-6785, (2020)
			
			\bibitem{Fine} Fine, B.: Classification of finite rings of order $p^2$. Mathematics magazine, \textbf{66}(4), 248-252, (1993).
			
			\bibitem{Gan_Sihme} Gan, C., Li, C., Mesnager, S., Qian, H.: On hulls of some primitive BCH codes and self-orthogonal codes. IEEE Trans. Inf. Theory, \textbf{67}(10), 6442-6455, (2021)
			
			\bibitem{BoundTable} Grassl, M.: Bounds on the minimum distance of linear codes \url{http://www.codetables.de}, Accessed on \today
			
			\bibitem{Griesmer} Griesmer, J.H.: A bound for error correcting codes, IBM J. Res. Dev., \textbf{4}(5), 532-542, (1960)
			
			\bibitem{Z4} Hammons, A.R., Kumar, P.V., Calderbank, A.R., Sloane, N.J., Sol\'e, P.: The $\mathbb{Z}_4$-linearity of Kerdock, Preparata, Goethals, and related codes. IEEE Trans. Inf. Theory \textbf{40}(2), 301-319, (1994)
			
			\bibitem{Hyun_Kim} Hyun, J.Y., Kim, H.K., Wu, Y., Yue, Q: Optimal minimal linear codes from posets. Des. Codes Cryptogr. \textbf{88}(12), 2475-2492, (2020)
			
			\bibitem{Hyun_Lee} Hyun, J.Y., Lee, J., Lee, Y.: Infinite families of optimal linear codes constructed from simplicial complexes. IEEE Trans. Inf. Theory \textbf{66}(11), 6762-6775, (2020)
			
						
			
			\bibitem{wchuffman} Huffman W.C., Pless, V.: Fundamentals of Error-Correcting Codes. Cambridge University Press, Cambridge, (2003)
			
			
			
			\bibitem{DNAnonunital} Kim, J.L., Ohk, D.E.: DNA codes over two noncommutative rings of order four. J. Appl. Math. Comput. \textbf{68}(3), 2015-2038, (2022)
			
			
			
			\bibitem{KimI} Kim, J.L., Roe, Y.G.: Construction of quasi self-dual codes over a commutative non-unital ring of order 4. Appl. Algebra Eng. Comm. Comput. 1-14, (2022), \url{https://doi.org/10.1007/s00200-022-00553-8} 
			
			
			\bibitem{Klove} Kl\o ve, T.: Codes for Error Detection, World Scientific, Singapore, (2007)
			
			\bibitem{GeneralCase} Liu, H., Yu, Z.: Linear Codes from Simplicial Complexes over $\mathbb {F} _ {2^ n} $, (2023) arXiv preprint arXiv:2303.09292, \url{https://arxiv.org/abs/2303.09292}
			
			\bibitem{F4} MacWilliams, F.J., Sloane, J.A.: The Theory of Error-Correcting Codes (North-Holland, Amsterdam), (1977)
					
			
			\bibitem{MasseyMinimal} Massey, J.L.: Minimal codewords and secret sharing, In Proc. 6th Joint Swedish-Russian Int. Workshop on Info. Theory, 276-279, 1993
			
			\bibitem{Sagar_Sarma} Sagar, V., Sarma, R.: Octanary linear codes using simplicial complexes. Cryptogr. Commun. (2022), \url{https://doi.org/10.1007/s12095-022-00617-z}
			
			\bibitem{Sagar_Sarma2} Sagar, V., Sarma, R.: Certain binary minimal codes constructed using simplicial complexes. (2022) arXiv preprint arXiv:2211.15747, \url{https://arxiv.org/abs/2211.15747}
			
			
			\bibitem{Sagar_Sarma4} Sagar, V., Sarma, R.: Codes over the non-unital non-commutative ring $E$ using simplicial complexes. (submitted)
			
			\bibitem{shi_x} Shi, M., Li, X.: A new family of optimal binary few-weight codes from simplicial complexes. IEEE Commun. Lett. \textbf{25}(4), 1048-1051, (2021)
			
			\bibitem{shi_nonchain} Shi, M., Li, X.: Few-weight codes over a non-chain ring associated with simplicial complexes and their distance optimal Gray image. Finite Fields Appl. \textbf{80}, 101994, (2022)
			
			
			
			\bibitem{shi_x2} Shi, M., Li, X.: New classes of binary few weight codes from trace codes over a chain ring. J. Appl. Math. Comput. \textbf{68}(3), 1869-1880, (2022)
			
			
			
			\bibitem{Shi_guan} Shi, M., Guan, Y., Wang, C., Sol\'e, P.: Few-weight codes from trace codes over $R_k$. Bulletin of the Australian Mathematical Society \textbf{98}(1), 167-174, (2018)
			
			
			
			\bibitem{shi_qian} Shi, M., Qian, L., Helleseth, T., Sol\'e, P.: Five-weight codes from three-valued correlation of M-sequences.
			Adv. Math. Commun., doi:10.3934/amc.2021022, (2021)
			
			
			\bibitem{MinjaNonunital} Shi, M., Wang, S., Kim, J.L., Sol\'e, P.: Self-orthogonal codes over a non-unital ring and combinatorial matrices. Designs, Codes and Cryptography, 1-13, (2021)
			
			
			\bibitem{shi_xuan} Shi, M., Xuan, W., Sol\'e, P.: Two families of two-weight codes over $\mathbb{F}_4$. Des. Codes Cryptogr. \textbf{88}(12), 2493-2505, (2020)
			
			\bibitem{LCDnonUnital} Shi, M., Li, S., Kim, J.L., Sol\'e, P.: LCD and ACD codes over a noncommutative non-unital ring with four elements. Cryptogr. Commun., 1-14, (2022)
			
			\bibitem{mixed2} Wang, S., Shi, M.: Few-weight ${\pmb {{\mathbb {Z}}}} _p\pmb {{\mathbb {Z}}} _p [u]$-additive codes from down-sets. J. Appl. Math. Comput. \textbf{68}(4), 2381-2388, (2022)	
			
			\bibitem{Wu_Lee} Wu, Y., Lee, Y.: Binary LCD codes and self-orthogonal codes via simplicial complexes. IEEE Commun. Lett., \textbf{24}(6), 1159-1162, (2020)
			
			\bibitem{Wu_Lee_2} Wu, Y., Lee, Y.: Self-orthogonal codes constructed from posets and their applications in quantum communication. Mathematics, \textbf{8}(9), 1495, (2020)
			
			\bibitem{Wu_Li} Wu, Y., Li, C., Xiao, F.: Quaternary linear codes and related binary subfield codes.  IEEE Trans. Inf. Theory \textbf{68}(5), 3070-3080, (2022)
			
			\bibitem{wu_zhu} Wu, Y., Zhu, X., Yue, Q.: Optimal few-weight codes from simplicial complexes. IEEE Trans. Inf. Theory \textbf{66}(6), 3657-3663, (2020)
			
			
			\bibitem{YuanDing} Yuan, J., Ding, C.: Secret sharing schemes from three classes of linear codes, IEEE Trans. Inf. Theory, \textbf{52}(1), 206-212, (2006)
			
			\bibitem{Zhou_Li} Zhou, Z., Li, X., Tang, C., Ding, C.: Binary LCD codes and self-orthogonal codes from a generic construction. IEEE Trans. Inf. Theory, \textbf{65}(1), 16-27, (2018)
			
			\bibitem{Zhu_Wei} Zhu, X., Wei, Y.: Few-weight quaternary codes via simplicial complexes. AIMS Math. \textbf{6}(5), 5124-5132, (2021)
			
		\end{thebibliography}
	\end{document}